\documentclass[aps,prl,superscriptaddress,showpacs, twocolumn]{revtex4}

\usepackage{graphicx}
\usepackage{epstopdf}
\usepackage{comment} 
\usepackage{amsmath}
\usepackage{amssymb}

\newcommand{\beq}{\begin{equation}}
\newcommand{\eeq}{\end{equation}}

\begin{document}

\title{Diverse forms of $\sigma$ bonding in two-dimensional Si allotropes: \textit{Nematic}  orbitals in the MoS$_2$ structure}

\author{Florian Gimbert}
\author{Chi-Cheng Lee}
\author{Rainer Friedlein}
\author{Antoine Fleurence}
\author{Yukiko Yamada-Takamura}
\affiliation{School of Materials Science, Japan Advanced Institute of Science and Technology (JAIST), 1-1 Asahidai, Nomi, Ishikawa 923-1292, Japan}
\author{Taisuke Ozaki}
\affiliation{School of Materials Science, Japan Advanced Institute of Science and Technology (JAIST), 1-1 Asahidai, Nomi, Ishikawa 923-1292, Japan}
\affiliation{Research Center for Simulation Science, Japan Advanced Institute of Science and Technology (JAIST), 1-1 Asahidai, Nomi, Ishikawa 923-1292, Japan}

\date{\today}

\begin{abstract}

The interplay of $sp^2$- and $sp^3$-type bonding defines silicon allotropes in two- and three-dimensional forms. 
A novel two-dimensional phase bearing structural resembleance to a single MoS$_2$ layer is found to possess a lower total energy than low-buckled silicene and to be stable in terms of its phonon dispersion relations. 
A new set of cigar-shaped, \textit{nematic} orbitals originating from the Si $sp^2$ orbitals realizes bonding with a 6-fold coordination of the inner Si atoms of the layer. 
The identification of these \textit{nematic} orbitals advocates diverse Si bonding configurations different from those of C atoms. 
\end{abstract}
\pacs{73.22.-f, 71.20.Mq, 61.48.-c}

\maketitle

With its forefront runner graphene and its unique and exotic properties, at present, two-dimensional materials experience an explosion of interest in scientific and technological aspects \cite{Geim07}. While the excellent electronic properties of graphene are derived from its structural robustness, the same property makes it a challenging task to engineer the optical and transport properties.
This challenge is stimulating the search for alternative two-dimensional layered materials 
that are more flexible in terms of its structural and electronic properties \cite{Cahangirov07, Fleurence12}.  
In this context, in particular, two new promising two-dimensional materials with a honeycomb structure made of silicon or germanium atoms have been studied as theoretical objects since 1994 \cite{Takeda94}. 
Most importantly, in yet hypothetical, slightly buckled, free-standing forms, silicene and germanene, as they were later called \cite{Guzman07}, exhibit a band structure similar to that of graphene, merging linear dispersion of $\pi$ and $\pi^{*}$ bands at the Fermi level to form Dirac cones at the $K$ points \cite{Takeda94, Guzman07, Cahangirov07}. 

Experimentally, it has been shown that two-dimensional Si honeycomb lattices can be formed epitaxially on Er layers \cite{Wetzel94} as well as on the Ag(111)\cite{Voigt12, Lin12, Jamgotchian12}, ZrB$_2$(0001)\cite{Fleurence12} and Ir(111)\cite{Meng13} surfaces. It is established that the interactions with the substrates have a distinct influence on the structural and electronic properties of the layers \cite{Chen12,Chen13,Lee13}. No experimental evidence for the existence of germanene has been reported yet.

As density functional theory (DFT) calculations predict consistently that slighly or low-buckled (LB) silicene 
is the most stable form of freestanding Si allotropes, very recently, it has been shown that the addition of Si adatoms on pristine silicene results in the formation of a dumbbell structure with a lower total energy \cite{Kaltsas13,Ongun13}. 
Interestingly, a higher cohesive energy can be achieved towards the complete coverage of the adatoms, which possesses even a higher cohesive energy than the low-buckled silicene. 
In fact, the periodic dumbbells can be recognized as the structure of a well-known single-layer of MoS$_2$. 
By considering the 4-fold coordination realized in the $sp^3$ bonding of Si atoms and also for the Si atom connecting the low-buckled silicene and a dumbbell, it is diffcult to understand why bonding with a 6-fold coordination could be formed by Si atoms in the MoS$_2$ structure. 
Therefore, it is timely and interesting to investigate the properties of this new Si phase beyond the view related to the introduction of defects or adatoms to low-buckled silicene.

In this Letter, by first-principles calculations \cite{Ozaki03,DFT}, we investigate  
the stability of this new Si phase (MoS$_2$-Si) together with a possible similar Ge allotrope (MoS$_2$-Ge), 
whose structures are that of a single layer of molybdenum disulfide, or MoS$_2$ \cite{Radisavljevic11}, 
in a wide range of lattice constants and compare it with that of other two-dimensional silicon structures. 
The phonon dispersion further demonstrates that MoS$_2$-Si is stable on the Born-Oppenheimer surface. 
To understand the 6-fold coordinated bonding in MoS$_2$-Si, we construct symmetry-respecting Wannier functions \cite{Vanderbilt,Weng}. 
A new form of $\sigma$ bonding expressed by three cigar-shaped orbitals co-exists with $\pi$ bands that are formed  by three additional reconstructed $p_z$ orbitals. 
The direction of these orbitals has changed from the typical in-plane direction of the $sp^2$ to the out-of-plane direction to form cigar-shaped orbitals.
In analogy to the nematic electronic structure \cite{Chuang}, the aligned and cigar-shaped orbitals may be called "\textit{nematic}". 
While as a common feature in the two-dimensional Si allotropes, the three $\sigma$ bands show dispersions similar to those of bands in low-buckled silicene, values of bond lengths and buckling heights vary significantly.  
Our finding suggests that the $\sigma$ bonds of Si atoms are more flexible than one could expect such that diverse forms of $\sigma$ bonding can allow the existence of a number of Si allotropes. 

\begin{figure}[ht!] 
\includegraphics[width=0.7\columnwidth]{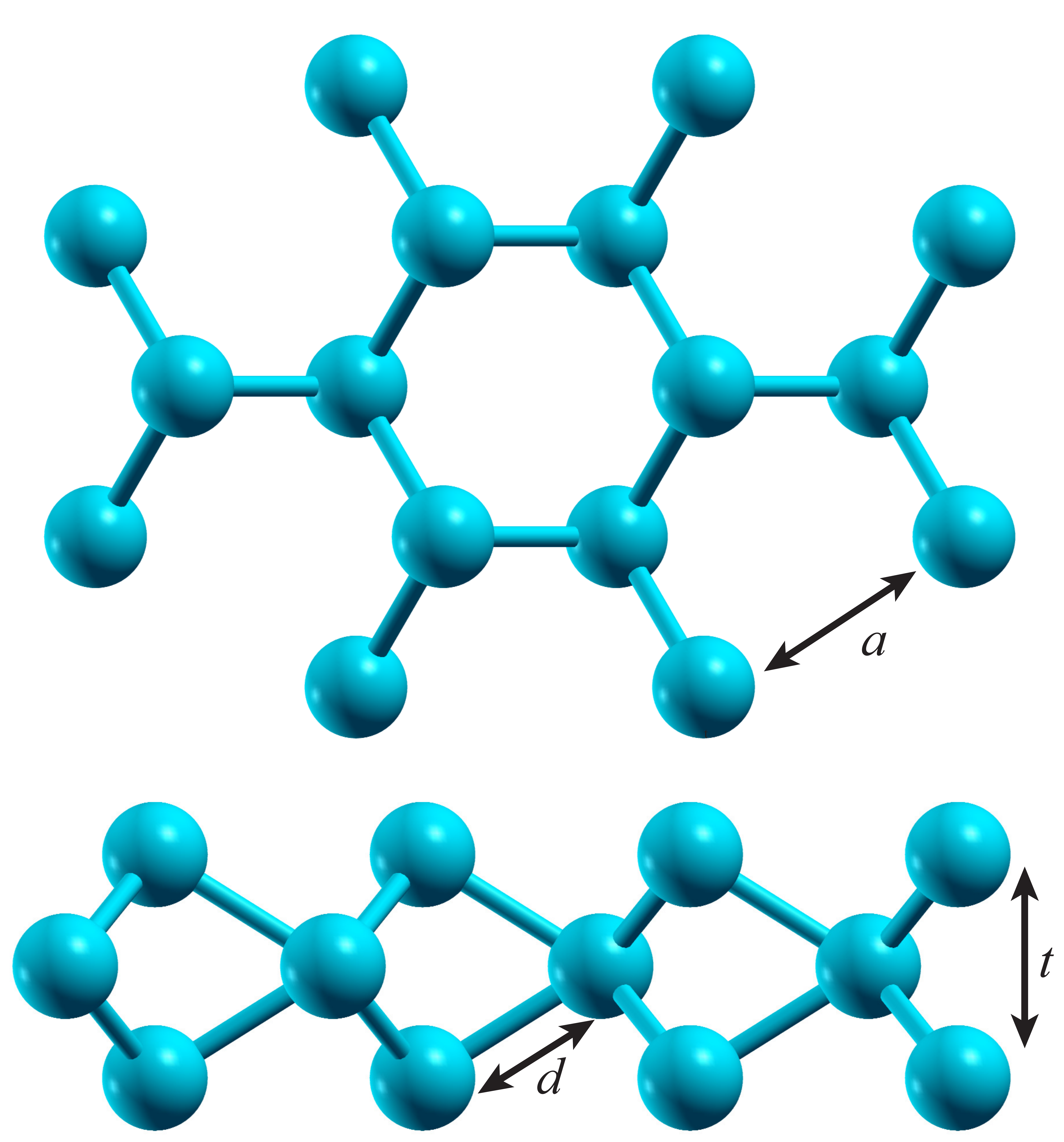}
\caption{Top and side view of the MoS$_2$-type single layer for silicon atoms with the lattice constant (a),  the bonding distance (d) and the thickness (t)  indicated.}
\label{Fig1}
\end{figure}

As shown in Figure 1, a single layer of MoS$_2$-Si crystallizes in an A-B-A stacking structure. 
In preserving the honeycomb structure, the A-B layer taken by itself is exactly freestanding low-buckled silicene.
In the MoS$_2$-Si structure, the primitive unit cell contains three atoms in comparison with two atoms in silicene. 
The middle atom is bound to six atoms while the top and bottom atoms have three neighbors. 
Clearly, the coordination is very different from that in graphene-like structures where each atom is bound to three neighboring atoms. 

\begin{table}[tbh]
\begin{tabular}{lcccc}
\hline
\hline
  & $a$ & $t$  & $d$ & $E_{r}$  \\
\hline
silicene & 3.90 & 0.49 &  2.30 & 0.032\\
MoS$_2$-Si & 3.64 & 2.63 &  2.47 & 0\\
\hline
germanene & 4.07 & 0.74 & 2.47 & 0.140\\
MoS$_2$-Ge & 3.90 & 2.88 & 2.67 & 0\\
\hline
\hline
\end{tabular}
\caption{Values of the lattice constant ($a$ in \AA), the thickness ($t$ in \AA), 
the bonding distance ($d$ in \AA) and the relative energy ($E_{r}$ in eV/atom) for silicene and MoS$_2$-Si (upper part), as well as germanene and MoS$_2$-Ge (lower part).}
\label{Tab}
\end{table}

\begin{figure}[ht!] 
\includegraphics[width=1\columnwidth]{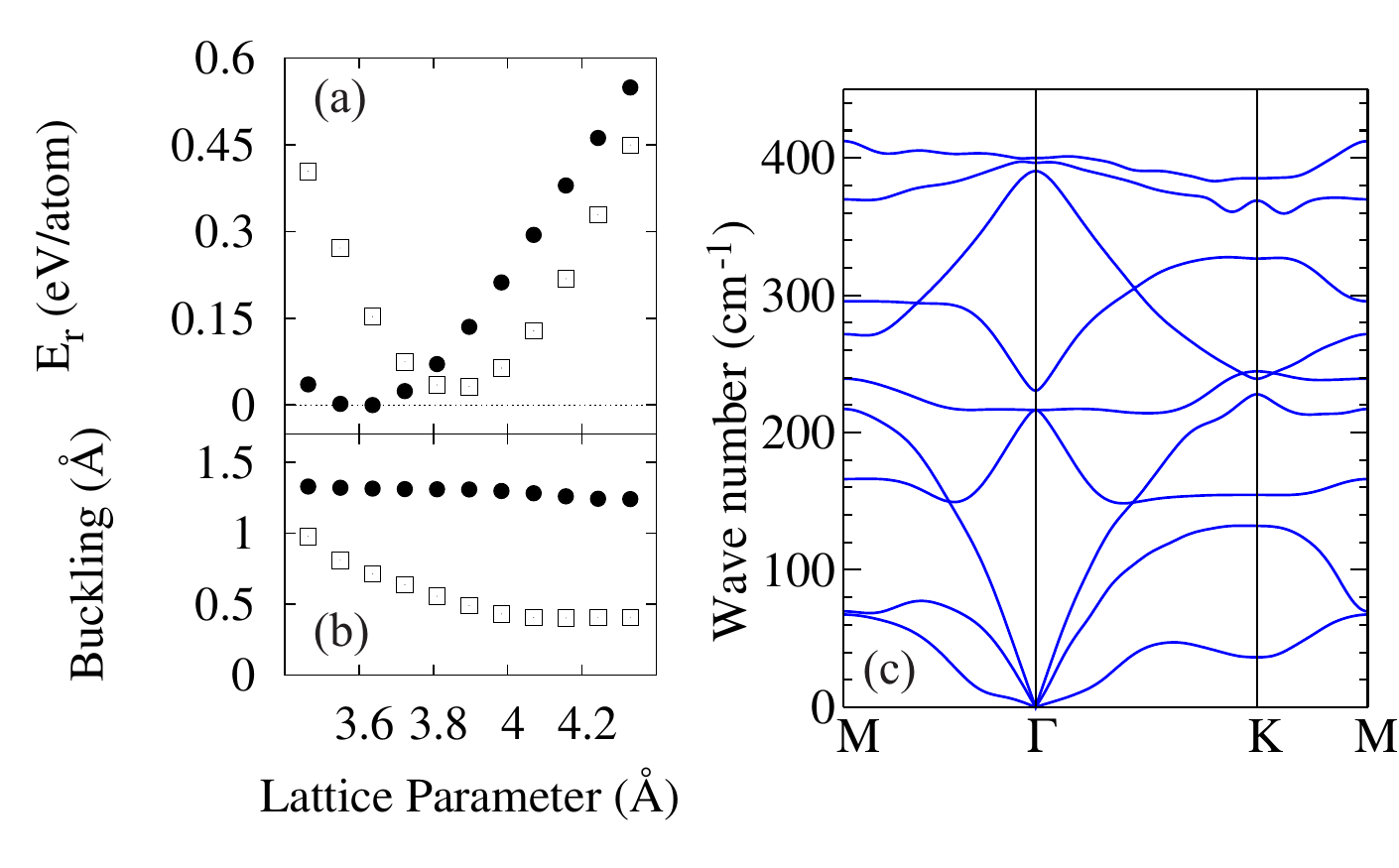}
\caption{(a): Relative energy (in eV/atom) of silicene (empty squares) and MoS$_2$-Si (filled circles). 
(b): Buckling of silicene (empty squares) and MoS$_2$-Si(filled circles). 
For MoS$_2$-Si, the buckling is defined as the distance beween two planes, corresponding to half of the thickness. 
(c): Phonon dispersion relations of MoS$_2$-Si as obtained by the force-constant method. }
\label{Fig2}
\end{figure}

Next, we compare the total energy per atom and structural parameters of the phases under consideration as a function of $a$. 
Since the high-buckled forms of silicene and germanene are unstable \cite{Cahangirov07}, 
we restrict the investigations to lattice parameters around to the region of the LB phase. In Fig. \ref{Fig2}(a) is shown the relationship between the total energy per atom $E_{r}$ and the lattice constant $a$, for both LB silicene (empty squares) and MoS$_2$-Si (filled circles). The evolution of the buckling is plotted in Fig. \ref{Fig2}(b). In order to facillitate comparison between the two phases, for the MoS$_2$ structure, half of the thickness is taken as the value of the buckling. 
With  the energy minimum occuring at a lattice constant of 3.90 \AA, the LB silicene prefers a buckling of 0.49 \AA\ and the energy minimum occurs at $a = 3.90$ \AA. These parameters compare well with values reported previously \cite{Cahangirov07, Houssa10, Wang13}. 

Interestingly, in equilibrium, the total energy of MoS$_2$-Si is lower than that of LB silicene, stabilized 
at a shorter lattice constant of $a=3.64$ \AA. Due to the stacking of three atoms, with 2.63 \AA, the thickness $t$ of the MoS$_2$-Si layer is larger than that of silicene.  
Similar to the Si phases, MoS$_2$-Ge possess a lower total energy than LB germanene as well. 
The total energy and lattice parameters of both MoS$_2$-Si and MoS$_2$-Ge are given in Tab.  \ref{Tab}. 

With MoS$_2$-Si being more stable than LB silicene, it is relevant to understand its stability by investigating the Born-Oppenheimer surface. This can be done by calculating the phonon frequencies in the harmonic approximation. 
In order to do so, a dynamical matrix has been constructed by calculating real-space force constants. 
Using a $12\times12$ supercell, the atoms have been chosen to be displaced by 0.02 \AA\ out of the equilibrium positions. 
The phonon dispersion relations of MoS$_2$-Si are shown in Fig. \ref{Fig2}(c). The frequencies are overall lower than those of LB silicene which can be understood from the elongated bond lengths that represent a weaker bonding. 
No branch with imaginary frequencies is found. 
This suggests that the freestanding MoS$_2$-Si is a stable phase that is preferentially formed instead of the commonly studied low-bucked silicene. 
For MoS$_2$-Ge, on the other hand, the situation is not as clear since a small amount of computed imaginary phonon frequencies maybe due to either a possible instability of the structure or artifact derived from numerical noise in the calculated forces.  
The following discussion will therefore focus on MoS$_2$-Si.

\begin{figure}[ht!] 
\includegraphics[width=1\columnwidth]{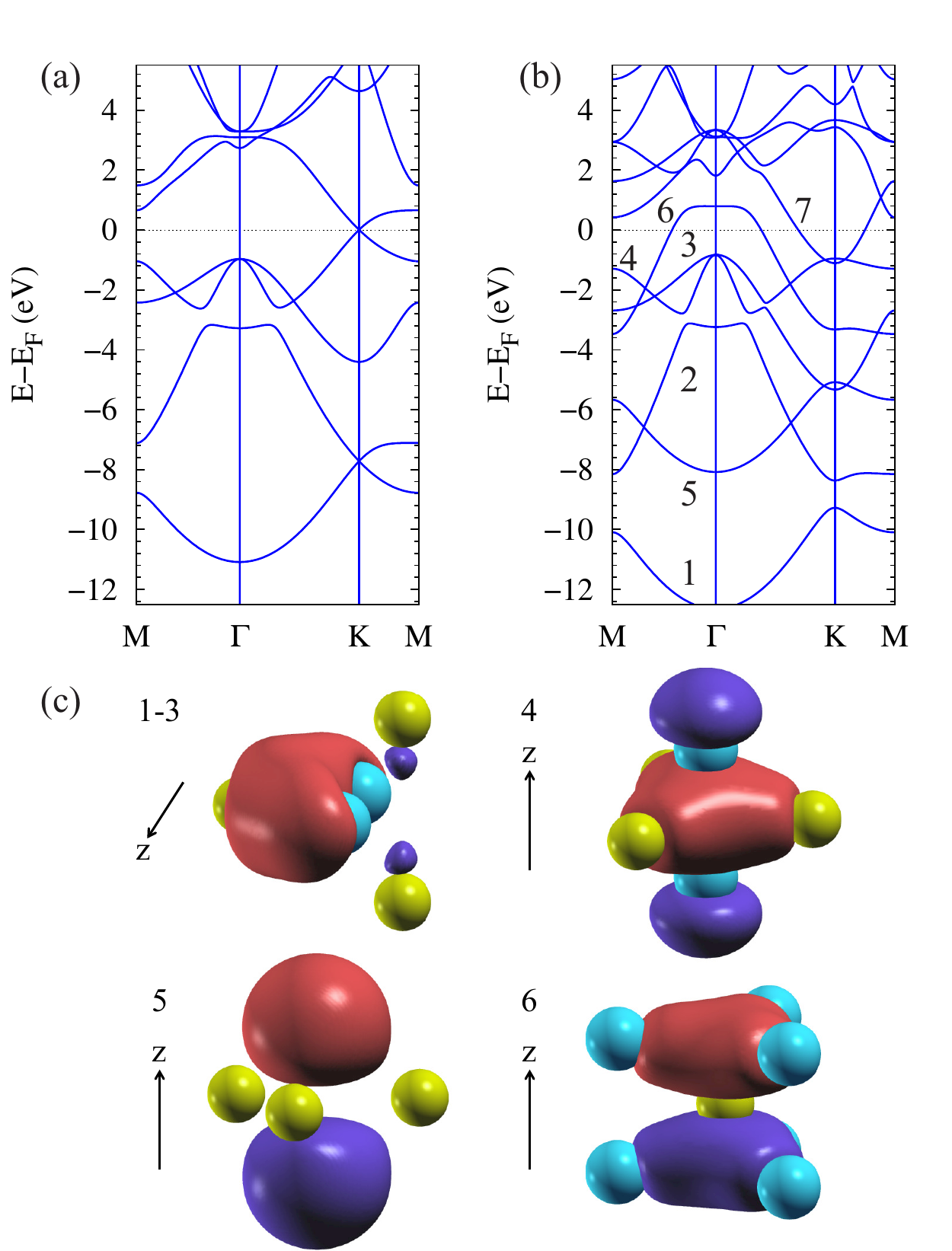}
\caption{(a): Band structure of low-buckled silicene  
with a lattice parameter of $a= 3.90$ \AA, 
(b): Band structure of MoS$_2$-Si with $a= 3.64$ \AA. 
(c): Symmetry-respecting Wannier functions of MoS$_2$-Si related to the bands labeled 1-6 in (b). 
The top, bottom and middle Si atoms are shown with different colors.}
\label{Fig3}
\end{figure}

The electronic band structure of LB silicene and MoS$_2$-Si are presented in Figs. \ref{Fig3}(a) and (b), respectively. 
Given that the bond length and the degree of buckling are larger and that the lattice constants are shorter for MoS$_2$-Si as compared to silicene, 
it is surprising that the band structure is not far away from that of silicene. 
Major differences observed around the Fermi energy relate to the disapearance 
of the Dirac cone at the K point typical for the LB silicene and to the appearance of two new bands, 
labeled 5 and 6 in Fig. \ref{Fig3}(b).

To further understand similarities and differences between these two phases, we construct the symmetry-respecting Wannier functions of MoS$_2$-Si. 
Technically, in order to do so, the  commonly adopted procedure for maximizing the localization of Wannier functions was not be performed \cite{Weng}.
The energy window is chosen to allow for a reproduction of the six occupied bands, labeled 1-6 in Fig. \ref{Fig3}(b).

As shown in Fig. \ref{Fig3}(c), the respective orbitals represented by Wannier functions have contributions in different bands and adopt particular shapes. 
The orbitals dominating the bands 1-3 originate from the $sp^2$ orbitals of the middle Si atom. 
Interestingly, in order to accomodate bonding between the top and bottom Si atoms, these orbitals have a \textit{nematic} shape. 
For clarity, only one of the three symmetric nematic orbitals is shown in Fig. \ref{Fig3}(c). The band dispersions related to the \textit{nematic} orbitals 
resemble those of the $\sigma$ bands of the LB silicene in Fig. \ref{Fig3}(a). Without these \textit{nematic} orbitals, it is difficult to provide \textit{fully occupied}
$\sigma$ band dispersions that are similar to the ones in LB silicene. 
Note that for the silicene structure of the A-B stacking obtained directly from the equilibrium lattice parameters of MoS$_2$-Si, the $\sigma$ bands of silicene can only be partially occupied.

Another interesting finding relates to orbitals with $p_z$ contributions: in particular, the 4$^{\rm th}$ orbital is derived from the $p_z$ orbitals of the top and bottom Si atoms and the $sp^2$ orbitals of the middle Si atoms. At the K point, the corresponding band crosses the 7$^{\rm th}$ band which, however, does not have any $p_z$ character. 
The crossing can therefore not be considered to be derived from an original Dirac point of silicene. 
The 5$^{\rm th}$ and 6$^{\rm th}$ orbitals form orbitals with $p_z$ symmetry having a node at the height of the middle Si atom. 
As it can be recognized in Fig. \ref{Fig3}(c), while the 5$^{\rm th}$ orbital is mainly derived from the $p_z$ and $s$ orbitals of the top and bottom Si atoms, the 6$^{\rm th}$ orbital stems from the $sp^2$ orbitals of the top and bottom Si atoms such displaying $p_z$ symmetry. 
Although the band dispersions of the $p_z$ orbitals bear some resemblance to the $\pi$ and $\pi^*$ bands of LB silicene, no Dirac cones are formed since the orbital nature of the two $p_z$ orbitals is essentially different.
 
\begin{figure}[ht!] 
\includegraphics[width=1\columnwidth]{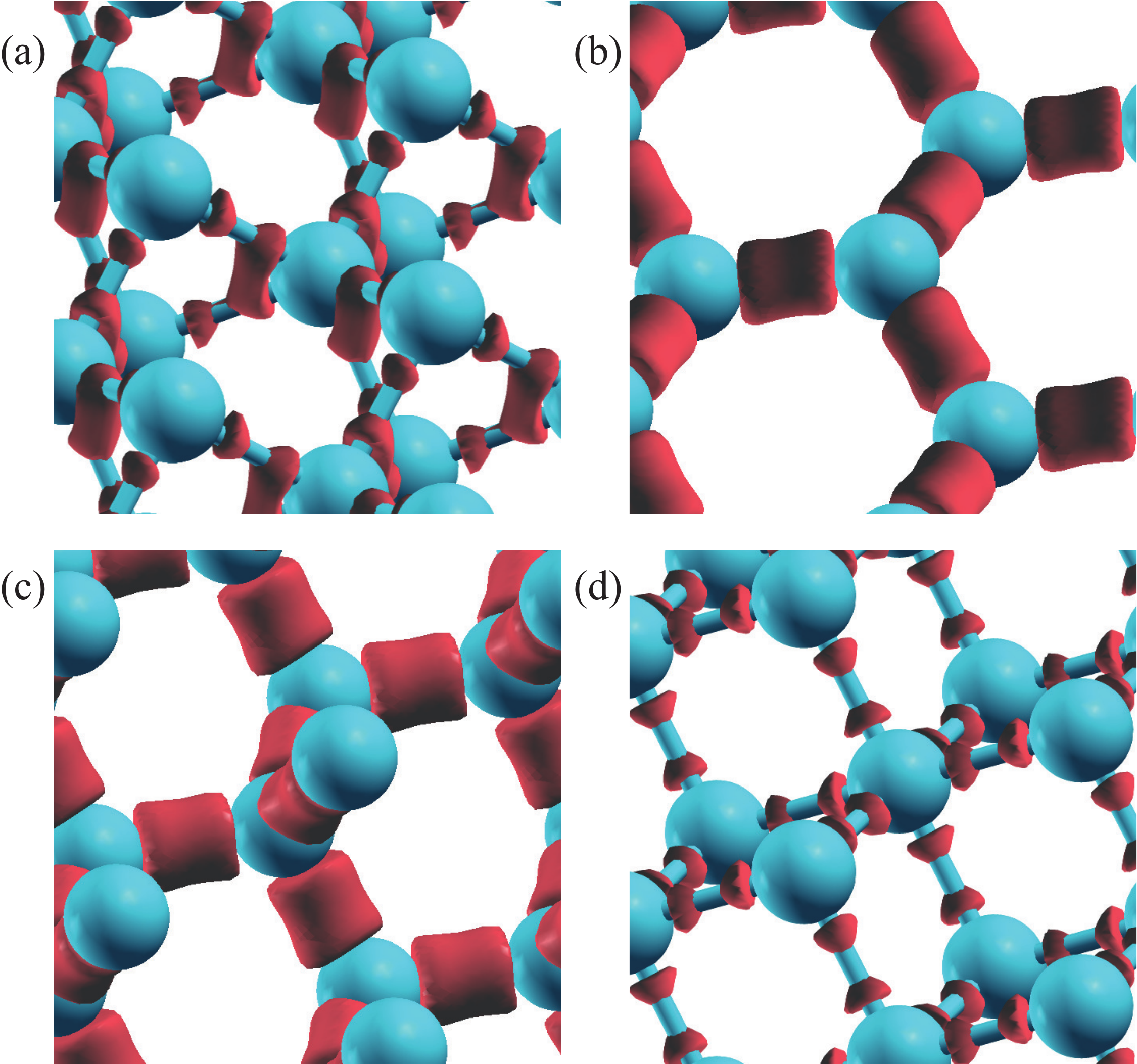}
\caption{Valence charge density for different structures composed of Si atoms. (a): silicon MoS$_2$-type  layer, corresponding to A-B-A stacking. (b): low-buckled silicene. (c): silicon layer with a diamond structure. (d): A-B-C stacking structure for silicon.}
\label{Fig4}
\end{figure}

The modification of the electronic properties is also evident from the plot of the charge density shown in Fig. \ref{Fig4}(a). The top and bottom atoms are bound to the central atom \textit{via} three of these \textit{nematic} orbitals which allow for the coordination of the central atoms with its six neighbors. 
For comparison, in Figs. \ref{Fig4}(b) and (c) are displayed the charge densities of silicene and of the corresponding Si layer in the diamond structure, in which atoms are 3- or 4-fold coordinated, respectively. Note that as single layers, these two structures have a higher total energy per atom than MoS$_2$-Si. This suggests that for two-dimensional silicon, $\sigma$ bonding of the planar $sp^2$ type is preferred. 
This symmetry is well respected by the \textit{nematic} orbital. 

In order to allow for an even higher flexibility in the bonding, one can imagine to twist the nematic orbitals to form a new A-B-C stacking structure. 
Such an asymmetric bonding with respect to the planar $sp^2$ orbitals is by 180 meV per Si atom energetically unfavourable.  
In addition, the middle Si atom of A-B-C stacking structure shares a bonding similar to that of the 6-fold coordinated Si atom in the $\beta$-tin Si phase that can only be stabilized under a high pressure \cite{Betatin}. The charge density of the A-B-C phase is shown in Fig. \ref{Fig4}(d).

To summarize, a new two-dimensional Si phase structurally equal to a single MoS$_2$ layer is identified.
Within DFT, this phase is stable on the Born-Oppenheimer surface in terms of the total energy and the phonon frequencies. 
Instead of the commonly accepted bonding configurations in silicene or the diamond structure of silicon that are related to a mixture of $sp^2$ and $sp^3$ or pure $sp^3$ hybridization, respectively, this phase exhibits \textit{nematic} orbitals that allow $\sigma$ bonding with a 6-fold coordination for the middle atoms in the MoS$_2$ structure. 
With bond lengths longer than for silicene, the three \textit{nematic} orbitals exhibit $\sigma$ band dispersions similar to those of LB silicene which represents a common feature of $\sigma$ bonding in two-dimensional Si phases. On the other hand, the reconstructed $p_z$ orbitals, or \textit{super} $p_z$ orbitals, are prominently different from those of low-buckled silicene. 
Per Si atom, the MoS$_2$-Si phase is lower in total energy than the low-buckled silicene making it the the most stable two-dimensional Si allotrope predicted so far. 
Our study demonstrates that Si atoms are capable of forming diverse types of $\sigma$ bonds even under  ambient  pressure conditions that are by themselves quite different from those formed by its smaller and larger cousins carbon and germanium. Even more, the presence of an extended $\pi$ electronic system with properties different to those in silicene and graphene will lead to properties that still must be explored. 
In a wider context, this finding does not only open new opportunities in the engineering of novel nanostructures to be employed in future applications but it also leads to intriguing questions for instance related to the Si-based chemistry.   

This work has been supported by the Strategic Programs for Innovative Research (SPIRE), MEXT, the Computational Materials Science Initiative (CMSI), by Materials Design through Computics: Complex Correlation and Non-Equilibrium Dynamics, A Grant in Aid for Scientific Research on Innovative Areas, MEXT, Japan, and by the Funding Program for Next Generation World-Leading Researchers (GR046).
The calculations have been performed using the Cray XC30 machine at Japan Advanced Institute of Science and Technology (JAIST).

\end{document}